%Paper: gr-qc/9403010
%From: STRING@TUHEP.PHY.TUFTS.EDU
%Date: Wed, 2 Mar 1994 15:35 EST

%\font\twelverm=cmr10	scaled 1200	\font\twelvei=cmmi10	scaled 1200
\font\twelverm=cmr12		\font\twelvei=cmmi12
\font\twelvesy=cmsy10	scaled 1200	\font\twelveex=cmex10	scaled 1200
\font\twelvebf=cmmib10	scaled 1200	\font\twelvesl=cmsl10	scaled 1200
%\font\twelvett=cmtt10	scaled 1200	\font\twelveit=cmti10	scaled 1200
\font\twelvett=cmtt12		\font\twelveit=cmti12
\font\twelvesc=cmcsc10	scaled 1200	\font\twelvesf=cmss10	scaled 1200
\font\twelvebfrm=cmbx10	scaled 1200
\font\tenbf=cmmib10
\skewchar\twelvei='177	\skewchar\twelvesy='60

%  Define \...point macros to change fonts and spacings consistently

	%	Melott
\def\twelvepoint{\normalbaselineskip=12.4pt plus 0.1pt minus 0.1pt
 \abovedisplayskip 12.4pt plus 3pt minus 9pt
 \belowdisplayskip 12.4pt plus 3pt minus 9pt
 \abovedisplayshortskip 0pt plus 3pt
 \belowdisplayshortskip 7.2pt plus 3pt minus 4pt
 \smallskipamount=3.6pt plus1.2pt minus1.2pt
 \medskipamount=7.2pt plus2.4pt minus2.4pt
 \bigskipamount=14.4pt plus4.8pt minus4.8pt
 \def\rm{\fam0\twelverm}		\def\it{\fam\itfam\twelveit}
 \def\sl{\fam\slfam\twelvesl}	\def\bf{\fam\bffam\twelvebf}
 \def\mit{\fam 1}		\def\cal{\fam 2}
 \def\sc{\twelvesc}		\def\tt{\twelvett}
 \def\sf{\twelvesf}		\def\bfrm{\twelvebfrm}
 \textfont0=\twelverm	\scriptfont0=\tenrm	\scriptscriptfont0=\sevenrm
 \textfont1=\twelvei	\scriptfont1=\teni	\scriptscriptfont1=\seveni
 \textfont2=\twelvesy	\scriptfont2=\tensy	\scriptscriptfont2=\sevensy
 \textfont3=\twelveex	\scriptfont3=\twelveex	\scriptscriptfont3=\twelveex
 \textfont\itfam=\twelveit	\textfont\slfam=\twelvesl
 \textfont\bffam=\twelvebf	\scriptfont\bffam=\tenbf
 \scriptscriptfont\bffam=\sevenbf
 \normalbaselines \rm}

%	tenpoint

%%	Various internal macros

\def\beginlinemode{\endmode \begingroup\parskip=0pt
	\obeylines\def\\{\par}\def\endmode{\par\endgroup}}
\def\beginparmode{\endmode \begingroup \def\endmode{\par\endgroup}}
\let\endmode=\par
{\obeylines\gdef\
{}}

	%	(for Melott)
	%	(for Melott)
\def\singlespace{\baselineskip=\normalbaselineskip}

\def\oneandahalfspace{\baselineskip=\normalbaselineskip
		\multiply\baselineskip by 3 \divide\baselineskip by 2}
\def\doublespace{\baselineskip=\normalbaselineskip \multiply\baselineskip by 2}

\newcount\firstpageno
\firstpageno=2
\footline={\ifnum\pageno<\firstpageno{\hfil}\else{\hfil\twelverm\folio\hfil}\fi}
\def\toppageno{ \global\footline={\hfil} \global\headline
  ={\ifnum\pageno<\firstpageno{\hfil}\else{\hfil\twelverm\folio\hfil}\fi}}
\let\rawfootnote=\footnote		% We must set the footnote style
\def\footnote#1#2{{\rm\singlespace\parindent=0pt\parskip=0pt
  \rawfootnote{#1}{#2\hfill\vrule height 0pt depth 6pt width 0pt}}}
\def\raggedcenter{\leftskip=4em plus 12em \rightskip=\leftskip
  \parindent=0pt \parfillskip=0pt \spaceskip=.3333em \xspaceskip=.5em
  \pretolerance=9999  \tolerance=9999
  \hyphenpenalty=9999 \exhyphenpenalty=9999 }
\def\dateline{\rightline{\ifcase\month\or
  January\or February\or March\or April\or May\or June\or
  July\or August\or September\or October\or November\or December\fi
  \space\number\year}}
\def\received{\vskip 3pt plus 0.2fill
  \centerline{\sl (Received\space\ifcase\month\or
  January\or February\or March\or April\or May\or June\or
  July\or August\or September\or October\or November\or December\fi
  \qquad, \number\year)}}

%%	Page layout, margins, font and spacing (feel free to change)

\hsize=6.5truein
\hoffset=0in	%	1truein
\vsize=9truein
\voffset=0in	%	1truein
\parskip=\medskipamount
\toppageno
\twelvepoint
\doublespace
\def\\{\cr}
\overfullrule=0pt % delete the nasty little black boxes for overfull box

%%	The user definitions for major parts of a paper (feel free to change)

  %  "Draft", Timestamp

	% Preprint number at upper right of title page

\def\title#1{			%  Title on title page
   \null \vskip 3pt plus 0.3fill \beginlinemode
   \doublespace \raggedcenter {\bfrm #1} \vskip 3pt plus 0.1 fill}

\def\author			%  Author(s) name(s)  on title page
  {\vskip 3pt plus 0.1fill \beginlinemode \doublespace \raggedcenter}

\def\affil			% Affiliations (can intermix with \author)
  {\vskip 3pt \beginlinemode \doublespace \raggedcenter \it}

\def\abstract			% Begin abstract
  {\vskip 3pt plus 0.1fill \subhead {Abstract:}
   \beginparmode \narrower \oneandahalfspace }

\def\endtopmatter		% End title page, begin body of paper
  {\vskip 3pt plus 0.1fill \endpage \body}

\def\body			% Begin text body;  can be used to end
  {\beginparmode}		% \title, \author, \affil, \abstract,
				% \reference, or \figurecaption modes

\def\head#1{			% Head;  NOTE enclose the text in {}
   \goodbreak \vskip 0.4truein	%  e.g., \head{I. Introduction}
  {\immediate\write16{#1} \raggedcenter {\sc #1} \par}
   \nobreak \vskip 3pt \nobreak}

\def\subhead#1{			% Subhead;  NOTE enclose the text in {}
  \vskip 0.25truein		% e.g., \subhead{A. History of the Problem}
  {\raggedcenter {\it #1} \par} \nobreak \vskip 3pt \nobreak}

\def\beneathrel#1\under#2{\mathrel{\mathop{#2}\limits_{#1}}}

\def\refto#1{${\,}^{#1}$}	% For references in text as superscript

\newdimen\refskip \refskip=0pt
% Begin references -- default style is Ap. J.,
%	i.e. [journal], [volume], [page] (space after comma).
\def\references{\head{References}
   \beginparmode \frenchspacing \parindent=0pt \leftskip=\refskip
   \parskip=0pt \everypar{\hangindent=20pt\hangafter=1}}

\gdef\refis#1{\item{#1.\ }}			% Ref list numbers.

\gdef\journal#1, #2, #3 {		% references,  Ap. J.  style
    {\it #1}, {\bfrm #2}, #3.}		% comma separates: name, vol, page

% NB: NYAcad puts year BEFORE journal name --
% ex:	J. Smith.  1901.  Gen. Rel. Grav.  123: 456.

%  ex:	J. Smith,  G. R. G.  123 (1901) 456.

% journal abbreviations: lower case ("\prd") is Phys. Rev. style

% uppercase ("\ApJ") for Ap. J. style

\def\endreferences{\body}

\def\figurecaptions		% Begin figure captions
  {\endpage \beginparmode \head{Figure Captions}
   \parskip=3pt \everypar{\hangindent=20pt\hangafter=1} }

\def\endpage{\vfill\eject}	%  Eject a page

% Ways to say goodbye
\def\endpaper	{\endmode\vfill\supereject}

\def\endjnl	{\endpaper\end}

%%	Various little user definitions

\def\ref#1{ref.{#1}}			%   For inline references
\def\Ref#1{Ref.{#1}}			% 	ditto
\def\[#1]{[\cite{#1}]}
\def\cite#1{{#1}}
		%   For citation of equation numbers
	%	ditto
			%	ditto

			%	ditto

%\def\Fig{Figure}
%\def\Figs{Figures}
		%   PRL figure caption page
		%   Ap. J. Figure caption page
\def\(#1){(\call{#1})}
\def\call#1{{#1}}
\def\frac#1#2{{#1 \over #2}}

\def\12{{1\over2}}

\def\sla{\raise.15ex\hbox{$/$}\kern-.57em}
\def\leaderfill{\leaders\hbox to 1em{\hss.\hss}\hfill}
\def\twiddle{\lower.9ex\rlap{$\kern-.1em\scriptstyle\sim$}}
\def\bigtwiddle{\lower1.ex\rlap{$\sim$}}

\def\gtwid{\mathrel{\raise.3ex\hbox{$>$\kern-.75em\lower1ex\hbox{$\sim$}}}}
\def\ltwid{\mathrel{\raise.3ex\hbox{$<$\kern-.75em\lower1ex\hbox{$\sim$}}}}
\def\square{\kern1pt\vbox{\hrule height 1.2pt\hbox{\vrule width 1.2pt\hskip 3pt
   \vbox{\vskip 6pt}\hskip 3pt\vrule width 0.6pt}\hrule height 0.6pt}\kern1pt}
\def\tdot#1{\mathord{\mathop{#1}\limits^{\kern2pt\ldots}}}

\def\pmb#1{\setbox0=\hbox{#1}%
  \kern-.025em\copy0\kern-\wd0
  \kern  .05em\copy0\kern-\wd0
  \kern-.025em\raise.0433em\box0 }

\catcode`@=11
\newcount\r@fcount \r@fcount=0
\newcount\r@fcurr
\immediate\newwrite\reffile
\newif\ifr@ffile\r@ffilefalse
\def\w@rnwrite#1{\ifr@ffile\immediate\write\reffile{#1}\fi\message{#1}}

\def\writer@f#1>>{}
\def\referencefile{%                      Stuff to write .REF file
  \r@ffiletrue\immediate\openout\reffile=\jobname.ref%
  \def\writer@f##1>>{\ifr@ffile\immediate\write\reffile%
    {\noexpand\refis{##1} = \csname r@fnum##1\endcsname = %
     \expandafter\expandafter\expandafter\strip@t\expandafter%
     \meaning\csname r@ftext\csname r@fnum##1\endcsname\endcsname}\fi}%
  \def\strip@t##1>>{}}

\def\citeall#1{\xdef#1##1{#1{\noexpand\cite{##1}}}}
\def\cite#1{\each@rg\citer@nge{#1}}     % Variable No. of args, separated by
%%","

\def\each@rg#1#2{{\let\thecsname=#1\expandafter\first@rg#2,\end,}}
\def\first@rg#1,{\thecsname{#1}\apply@rg}       % each@ag is a general purpose
\def\apply@rg#1,{\ifx\end#1\let\next=\relax%      variable no. of arg. macro.
\else,\thecsname{#1}\let\next=\apply@rg\fi\next}% args separated by commas

\def\citer@nge#1{\citedor@nge#1-\end-}  % Check for M-N range (M and N numbers)
\def\citer@ngeat#1\end-{#1}
\def\citedor@nge#1-#2-{\ifx\end#2\r@featspace#1 % Single argument
  \else\citel@@p{#1}{#2}\citer@ngeat\fi}        % M-N range of arguments
\def\citel@@p#1#2{\ifnum#1>#2{\errmessage{Reference range #1-#2\space is bad.}
    \errhelp{If you cite a series of references by the notation M-N, then M and
    N must be integers, and N must be greater than or equal to M.}}\else%
 {\count0=#1\count1=#2\advance\count1
by1\relax\expandafter\r@fcite\the\count0,%

  \loop\advance\count0 by1\relax%         Loop from M to N
    \ifnum\count0<\count1,\expandafter\r@fcite\the\count0,%
  \repeat}\fi}

\def\r@featspace#1#2 {\r@fcite#1#2,}    % Eat spaces at beginning or end of arg
\def\r@fcite#1,{\ifuncit@d{#1}          % Cite individual reference
    \expandafter\gdef\csname r@ftext\number\r@fcount\endcsname%
    {\message{Reference #1 to be supplied.}\writer@f#1>>#1 to be supplied.\par
     }\fi%
  \csname r@fnum#1\endcsname}

\def\ifuncit@d#1{\expandafter\ifx\csname r@fnum#1\endcsname\relax%
\global\advance\r@fcount by1%
\expandafter\xdef\csname r@fnum#1\endcsname{\number\r@fcount}}

\let\r@fis=\refis                       % Save old \refis, redefine
\def\refis#1#2#3\par{\ifuncit@d{#1}%      Use two params #2 #3 to strip blank
    \w@rnwrite{Reference #1=\number\r@fcount\space is not cited up to now.}\fi%
  \expandafter\gdef\csname r@ftext\csname r@fnum#1\endcsname\endcsname%
  {\writer@f#1>>#2#3\par}}

\def\r@ferr{\endreferences\errmessage{I was expecting to see
\noexpand\endreferences before now;  I have inserted it here.}}
\let\r@ferences=\references
\def\references{\r@ferences\def\endmode{\r@ferr\par\endgroup}}

\let\endr@ferences=\endreferences
\def\endreferences{\r@fcurr=0%            Save old \endreferences, redefine
  {\loop\ifnum\r@fcurr<\r@fcount%         Loop over refnum and produce text
    \advance\r@fcurr by 1\relax\expandafter\r@fis\expandafter{\number\r@fcurr}%
    \csname r@ftext\number\r@fcurr\endcsname%
  \repeat}\gdef\r@ferr{}\endr@ferences}

% Save old \endpaper, redefine it to write parting message.

\let\r@fend=\endpaper\gdef\endpaper{\ifr@ffile
\immediate\write16{Cross References written on []\jobname.REF.}\fi\r@fend}

\catcode`@=12

\citeall\refto          % These macros will generate citations
\citeall\ref            %
\citeall\Ref            %

%%%%%%%%%%%%%%%%%%%%%%%%%%%%%%%%%%%%%%%

\doublespace
\vglue 0. truein
\title
{
Approaches to Quantum Cosmology
}
\smallskip
\author
{Alexander Vilenkin}
\affil
{
Tufts Institute of Cosmology, Department of Physics and Astronomy,
Tufts University, Medford, MA 02155.
}

\abstract
\doublespace

Different proposals for the wave function of the universe are analyzed, with an
emphasis on various forms of the tunneling proposal.  The issues discussed
include the equivalence of the Lorentzian path integral and outgoing - wave
proposals, the definitions of the outgoing waves and of superspace boundaries,
topology change and the corresponding modification of the Wheeler - De Witt
equation.  Also discussed are the "generic" boundary condition and the third
quantization approach.

\endtopmatter
%\body

% to refer: \refto{dewitt} (like a superscript)
% to cite: \cite{dewitt} (eg. Ref. ...)

\head{1. Introduction}

In quantum
cosmology the whole universe is treated quantum-mechanically and is described
by a wave function rather than by a classical spacetime.  This quantum approach
to cosmology was originated by DeWitt\refto{dewitt} more than 25 years ago,
and after a
somewhat slow start has attracted much interest during the last decade.  The
picture that has emerged from this line of development\refto{tryon, fomin,
atpag,
av82, grizel, hh, linde, rub84, av84} is that a small closed universe can
spontaneously nucleate out of nothing, where "nothing" refers to the absence of
not only matter, but also of space and time.

The wave function of the universe is defined on superspace, which is the space
of all 3-metrics
$h_{ij}(x)$ and matter field configurations $\phi (x)$,
$$
\psi [h_{ij}(x), \phi (x)]   \,.          \eqno(1.1)
$$
It is invariant under 3-dimensional diffeomorphisms and satisfies the
Wheeler-DeWitt equation [1]
$$
{\cal H} \psi [h_{ij}, \phi] =0 \,.       \eqno(1.2)
$$
Here, ${\cal H}$ is a second-order differential operator in superspace.  In
principle, $\psi (h, \phi)$ should contain the answers to all meaningful,
questions one can ask about the universe.  However, the conditions necessary to
specify the appropriate solution of eq.(1.2) and the procedure by which
information can be extracted from that solution are far from being understood.

As (almost) any differential equation, the Wheeler-DeWitt equation has an
infinite number of solutions.  To get a unique solution, one has to specify
some boundary conditions in superspace.  In ordinary quantum mechanics, the
boundary conditions for the wave function are determined by the physical setup
external to the system under consideration.  In quantum cosmology,
there is nothing external to the universe, and it appears that a boundary
condition should be added to eq. (1.2) as an independent
 physical law.

Several candidates for this law of boundary condition have been proposed.
Hartle and Hawking\refto{hh} suggested that $\psi (h, \phi)$ should be given by
a
Euclidean path integral over compact 4-geometries $g_{\mu\nu}
(x, \tau)$ bounded by the 3-geometry $h_{ij}(x)$ with the field configuration
$\phi (x)$,
$$
\psi = \int^{(h, \phi)} [dg][d\phi] \exp [-S_E (g, \phi)] \,. \eqno(1.3)
$$
In this path-integral representation, the boundary condition corresponds to
specifying the class of histories integrated over in eq.(1.3).  Compact
4-geometries can be thought of as histories interpolating between a point
("nothing") and a finite 3-geometry $h_{ij}$.

A Euclidean rotation of the time axis, $t \to -i\tau$, is often used
in quantum field theory because it improves the convergence of the path
integrals.  However, in quantum gravity the situation is
 the opposite.  The gravitational part of the Euclidean action $S_E$ is
unbounded from below, and the integral (1.3) is badly divergent.  Attempts to
fix this problem by analytic continuation\refto{ghp} were only partly
successful,
and at present it remains unclear whether one can meaningfully define an
integral such as (1.3).

Alternatively, I proposed\refto{av84, farhi} that $\psi (h, \phi)$ should be
obtained by integrating over Lorentzian
histories interpolating between a vanishing 3-geometry ${\emptyset}$ and
 $(h, \phi)$ and lying to the past of $(h, \phi)$,
$$
\psi (h, \phi) = \int_{\emptyset}^{(h, \phi)} [dg][d\phi] e^{iS} \,. \eqno(1.4)
$$
This wave function is closely related to Teitelboim's causal propagator
\refto{teitel80, teitel82}
$K(h_2 ,\phi_2 | h_1, \phi_1)$,
$$
\psi (h, \phi) = K(h, \phi |{\emptyset})  \,. \eqno(1.5)
$$

Linde\refto{linde} suggested that, instead of the standard Euclidean rotation
$t \to
-i\tau$, the action $S_E$ in (1.3) should be obtained by rotating
 in the opposite sense, $t \to +i\tau$.  This gives a convergent path integral
for the scale factor, which is all one needs in the simplest
minisuperspace models.  But in models including matter degrees of freedom or
inhomogeneous modes of the metric one gets a divergent integral.  Additional
contour rotations might fix this problem, but no specific proposals have yet
been formulated.  Halliwell and Hartle\refto{halhar} discussed a path integral
over
complex metrics which are not necessarily purely Lorentzian or purely
Euclidean.  This encompases all of the above proposals and opens new
possibilities.  However, the space of complex metrics is very large, and no
obvious choice of the integration contour suggests itself as the preferred one.

In addition to these path-integral no-boundary proposals, one candidate law of
boundary conditions has been formulated directly as a boundary condition in
superspace.  This is the so-called tunneling boundary condition\refto{av86,
av88} which requires that $\psi$ should include only outgoing waves at
boundaries of superspace.  The main weakness of this proposal is that "outgoing
waves" and the "boundary of superspace" have not been rigorously defined.  The
Lorentzian path-integral proposal (1.4) was originally suggested\refto{av84} as
a path integral version of the tunneling boundary condition,
and indeed the two proposals give the same
 wave function in the simplest minisuperspace model\refto{hallou89}.  In the
general
case, the equivalence
 of the two proposals is far from being obvious.

I should also mention a completely different approach to quantum cosmology, the
so-called third
quantization\refto{rub88, mcgui, hosmor, gistro, banks88, fkps}.  Here, the
wave function of the universe $\psi$ is promoted to a quantum field operator
and is expressed in terms of creation and annihilation operators for the
universes.  The problem of defining the boundary conditions is then replaced by
the problem of determining the in-state of the quantum field $\psi$.  With the
radius of the universe playing the role of time, it is argued that creation of
universes from nothing corresponds to an "in-vacuum" state at vanishing radius.

The problem
of boundary conditions for the cosmological wave function is related to the
problem of topology change in quantum gravity.  In the path integral approach,
one has to specify whether the integration in (1.3), (1.4) is performed over
4-manifolds of arbitrary topology, or only a restricted class of topologies is
included.  In the tunneling approach, part of superspace boundary corresponds
to boundaries between different topological sectors, and one has to decide what
kind of boundary condition should be imposed there.  Moreover, we shall see in
Section 7 that topology change not only affects the boundary conditions for
$\psi$, but also leads to a modification of the Wheeler-DeWitt equation.

In this paper I shall review the status of the tunneling wave function of the
universe and attempt a more precise
formulation of the tunneling boundary condition.  As a prototype for this
boundary condition, the next section discusses the process of bubble nucleation
 in a false vacuum, which is in many ways analogous to the nucleation of
universes.  The outgoing-wave boundary condition for a nucleating bubble will
be formulated using a spherical minisuperspace model.  In Section 3, similar
approach is applied to the simplest cosmological minisuperspace model: a
Robertson-Walker universe with a cosmological constant,
$\Lambda > 0$.  In Section 4, the wave function for the same model is obtained
by analytic continuation from the "bound-state" wave function for $\Lambda <
0$.  Section 5 discusses the Lorentzian path integral approach and its
equivalence to the outgoing wave boundary condition.  A possible extension of
these approaches beyond minisuperspace is discussed in Section 6.  There, it is
suggested that some general properties of the potential term in the
Wheeler-DeWitt equation may allow one to define outgoing waves in the general
case.

The issues of topology change is tackled in Section 7.  It is argued that
topology-changing transitions can occur through superspace boundaries, but
generally involve configurations in the interior of superspace.  This implies
that the Wheeler-DeWitt equation needs to be modified.  A possible form of this
modified equation is suggested.  Section 8 gives some critical comments on the
third-quantization approach to topology change, and Section 9 contains some
concluding remarks.

\head{2. Bubble Nucleation}

To discuss the nucleation of true vacuum bubbles in a metastable false
vacuum, we shall make a number of simplifying assumptions.  First, we shall
assume that the bubble radius at nucleation is much greater than the thickness
of the bubble wall, so that the bubble can be approximated as an infinitely
thin sheet.  Second, we shall use the semiclassical approximation, assuming
that the tunneling action is large.  (This is always true for a thin-wall
bubble, provided the theory is weakly interacting).  The nucleating bubble is
then nearly spherical and can be adequately
described by a minisuperspace model with a single degree of freedom, the
bubble radius $R$.  Finally, we shall disregard the gravitational effects of
the false vacuum and assume the spacetime to be
Minkowskian.

In our minisuperspace model, the worldsheet
of the bubble wall is described by a single function $R(t)$, and the Lagrangian
 is
$$
L = -4\pi \sigma R^2 (1-{\dot R}^2)^{1/2} + {4\pi \over {3}}\epsilon R^3 \,.
\eqno(2.1)
$$
Here, $\sigma$ is the wall tension, $\epsilon$ is the
 difference between the energy densities of the false and true vacuum, and
${\dot R} = dR/dt$.  The momentum conjugate to the variable $R$ is
$$
p_R = 4\pi \sigma R^2 {\dot R} (1 - {\dot R}^2 )^{-1/2} \, \eqno(2.2)
$$
and the Hamiltonian is
$$
{\cal H} = [p_R^2 + (4\pi \sigma R^2)^2]^{1/2} - {4\pi \over{3}}\epsilon R^3
\,. \eqno(2.3)
$$

Bubble nucleation does not change the energy of the system, and if the false
vacuum energy
is set equal to zero, we have
$$
{\cal H}=0 \,, \eqno(2.4)
$$
which can be rewritten using (2.3),
$$
p_R^2 + U(R) = 0 \,, \eqno(2.5)
$$
$$
U(R) = (4\pi \sigma R^2)^2 (1 - R^2 / R_0^2) \,, \eqno(2.6)
$$
where $R_0 = 3\sigma /\epsilon$.  The equation of motion for $R(t)$ can be
obtained
from (2.2), (2.5) and (2.6),
$$
R^2 {\dot R}^2 = R^2 - R_0^2 \,, \eqno(2.7)
$$
and the solution is
$$
R(t) = (R_0^2 + t^2)^{1/2}  \,.  \eqno(2.8)
$$

The worldsheet metric of the bubble is
$$
ds^2 = (1-{\dot R}^2 )dt^2 - R^2 (t)d\Omega^2 \,, \eqno(2.9)
$$
where $d\Omega^2$ is the metric on a unit sphere.  With a new time coordinate,
$$
\tau = R_0 \sinh^{-1}(t/R_0) \,, \eqno(2.10)
$$
we recognize it as the metric of a (2+1)-dimensional de Sitter space,
$$
ds^2 = d\tau^2 - R^2(\tau)d\Omega^2 \,,
$$
$$
R(\tau) = R_0 \cosh (\tau / R_0) \,.  \eqno(2.11)
$$
If the bubble wall gets inhabited by some 2-dimensional creatures, they will
find themselves
 living in an expanding inflationary universe.  If they are smart enough, they
may also figure out that their universe was spontaneously created at
$\tau = 0$, and thus eq.(2.11) applies only for $\tau >0$.

How would these 2-dimensional physicists describe the
 quantum nucleation of the universe?  In quantum theory, the energy
conservation (2.4) gets replaced by
$$
{\cal H}\psi = 0 \,, \eqno(2.12)
$$
where $\psi (R)$ is the "wave function of the universe" and the momentum
operator is $p_R = -i\partial /\partial R$.  The square root in (2.3) is
complicated to deal with, and it is much easier to use the energy conservation
law in the form (2.5),
$$
[-\partial_R^2 + U(R)]\psi = 0 \,. \eqno(2.13)
$$
The transition from (2.12) to (2.13) involves commutation of
the non-commuting operators $R$ and $p_R$, which is justified, as long as
$$
|Rp_R \psi | >> |[R, P_R ]\psi | = |\psi| \,, \eqno(2.14)
$$
that is, away from the classical turning points, where
$p_R \approx 0$.  Using the classical equations of motion for $R(t)$, we find
that (2.14) is violated in a small neighborhood of the turning point $R_0$,
$$
\delta R/R_0 \sim (\sigma R_0^3)^{-2} << 1  \,. \eqno(2.15)
$$
Since the correct operator ordering is not known, we shall keep the simplest
choice as in (2.13).

We now come to the problem of determining the boundary
conditions for $\psi (R)$.  Only one non-trivial condition is required; the
second would simply determine the overall multiplicative constant.  The WKB
solutions of eq.(2.13) for $R > R_0$ are
$$
\psi_{\pm} (R) = p(R)^{-1/2} \exp \left(\pm i\int_{R_0}^R p(R')dR' \mp
i\pi /4 \right) \,,
\eqno(2.16)
$$
where
$$
p(R) = [-U(R)]^{1/2} \, \eqno(2.17)
$$
is the classical momentum.  To the leading order in the WKB approximation,
$$
{\hat p}_R \psi_{\pm} (R) \approx \pm p(R) \psi_{\pm} (R) \,, \eqno(2.18)
$$
where${\hat p}_R = -i\partial /\partial R$.  This shows that $\psi_+ (R)$ and
$\psi_- (R)$ describe, respectively, the expanding and contracting bubbles.
In the quantum nucleation process, only an expanding bubble must be present,
and thus we require that for $R > R_0$ the wave function should include only
the outgoing wave, $\psi_+ (R)$.

In the classically forbidden range, $0<R<R_0$, the two
 solutions of (2.13) are
$$
{\tilde \psi}_{\pm} = |p(R)|^{-1/2} \exp \left(\pm \int_R^{R_0} |p(R)|dR
\right) \,.
\eqno(2.19)
$$
With the outgoing wave boundary condition at large R, the wave function in this
 range is determined\refto{llqm}
 by matching at $R \approx R_0$,
$$
\psi (R<R_0) = {\tilde \psi}_+ (R) + {i \over{2}} {\tilde \psi}_- (R) \,.
\eqno(2.20)
$$
The two terms on the right-hand side of (2.20) have comparable magnitude
at $R \approx R_0$, but in the most of the forbidden range the ${\tilde \psi}_+
(R)$ term dominates.  The exponential factor in the tunneling probability can
be
 determined\refto{vko, col, calcol} from
$$
|{\psi (R_0) \over {\psi (0)}}|^2 \sim \exp \left(-2 \int_0^{R_0} |p(R)| dR
\right) = \exp (-\pi^2 \sigma R_0^3 /2) \,. \eqno(2.21)
$$
A different choice of operator ordering would not affect (2.21), but it could
affect the pre-exponential coefficient.

Having obtained the result (2.21), the 2-dimensional physicist could be puzzled
about its meaning
  What does it mean to find the nucleation probability for a bubble when there
is only one bubble?  Even if we assume that there
 are other bubbles, they are unobservable, so how can we test this theory
observationally?  Of course, in the case of a nucleating bubble, there is an
external observer for whom the nucleation probability has a well-defined
meaning.  This may or may not be
 so in the case of the universe.  But the point I want to make is that even a
worldsheet observer can derive some useful information from the wave function
of the universe.  If, for example, several different types of bubble can
nucleate, with different values of $\sigma$ and $\epsilon$, then the observer
is more likely to find herself in the type of bubble with the highest
nucleation probability (assuming, of course, that such bubbles are suitable for
 2-d life).

Furthermore, nucleating bubbles are not exactly
spherical, and one could in principle calculate the amplitude for a bubble to
have a given shape.  This problem has been solved in the perturbative
superspace approximation which includes all the degrees of freedom of the
bubble, but treats all but radial
 motions as small perturbations\refto{vv91, gv92}.  It turns out that
perturbations of a
spherical bubble can be represented as excitations
of a scalar field $\Phi$ that lives on the bubble worldsheet and has a
tachyonic mass, $m^2 = -3R_0^2$.  The mode expansion of this field contains
four "zero modes" which represent overall space and time translations of the
bubble, while the remaining modes describe deviations from spherical shape.
As in the cosmological case\refto{halhaw, vv88}, one finds that the bubble
nucleates with the
field $\Phi$ in a de Sitter-invariant quantum state.\refto{foot1}
This prediction should be testable both by external and worldsheet observers.

Extension of this analysis beyond perturbative superspace is a very
complicated problem which has not yet been solved. The bubble worldsheet can,
in general, be represented in a parametric form
 as $x^\mu (\xi^a)$ with $a=0,1,2$.  An external observer would evaluate the
amplitude to find a bubble in a given configuration at $x^0 = T$ by
evaluating the path integral
$$
\psi = \int [dx^\mu ] e^{iS} \,. \eqno(2.22)
$$
Given that there was no bubble at $x^0 = 0$, the integration should be taken
over all compact worldsheets bounded by the given 2-surface at $x^0 = T$ and
satisfying $0<x^0 (\xi)<T$.  As I said, calculating the integral (2.22), or
even making it well defined, is a very difficult problem.

Fot a worldsheet observer, $\xi^0$ is a time coordinate and $x^\mu (\xi)$ is a
set of four interacting scalar
fields.  She would find the restriction on the range of $x^0 (\xi)$ unnatural
and would probably define the no-boundary wave function
 $\psi$ in (2.22) as an unrestricted integral over $x^\mu (\xi)$.  The two
wave functions will generally be different, but in the semiclassical
 regime the integral (2.22) is dominated by the neighborhood of the classical
path, and the wave functions will be essentially the same.  It would be
interesting to further investigate this connection between the bubble wave
functions from worldsheet and target space points of view.  At present,
eq.(2.22) is purely formal, and its connection to the standard Euclidean
formalism\refto{col, calcol} for calculating the vacuum decay rate is obscure.

On a qualitative level, one expects quantum fluctuations to grow
large at small length scales, and if large deformations are allowed, then the
bubble wall can cross itself, and small "daughter
bubbles" can be chopped off.  When viewed at very small scales, the bubble wall
 may in fact have a fractal structure, with a dense
 foam of small bubbles surrounding it.  Moreover, the worldsheet observer may
discover that on sufficiently small scales her
bubble is not a 2-d surface after all, but is more adequately described by
certain solutions of (3+1)-dimensional field equations.
 Similar problems may face human observers as they explore distances
approaching the Planck scale.

\head{3. de Sitter Minisuperspace}

Turning now to the cosmological wave function, we first consider the simplest
minisuperspace model,
$$
S = \int d^4 x \sqrt{-g} \left( {R \over{16\pi G}} -\rho_v \right) \,,
\eqno(3.1)
$$
where $\rho_v$
is a constant vacuum energy and the universe is assumed to be homogeneous,
isotropic, and closed:
$$
ds^2 = \sigma^2 [N^2 (t) dt^2 - a^2 (t)d\Omega_3^2 ] \,. \eqno(3.2)
$$
Here, $N(t)$ is an arbitrary lapse function, $d\Omega_3^2$ is the metric on a
unit 3-sphere, and $\sigma^2 = 2G/3\pi$ is a normalizing factor chosen for
later
convenience.  Substituting (3.2) into (3.1), we obtain the Lagrangian
$$
{\cal L} = {1 \over{2}}N \left[
a \left( 1-{{\dot a}^2 \over{N^2}} \right) -\Lambda
a^3 \right] \,, \eqno(3.3)
$$
and the momentum
$$
p_a = -a{\dot a}/N \,, \eqno(3.4)
$$
where $\Lambda = (4G/3)^2 \rho_v$.
The Lagrangian (3.3) can also be expressed in the canonical form,
$$
{\cal L} = p_a {\dot a} - N{\cal H} \,, \eqno(3.5)
$$
where
$$
{\cal H} = -{
1 \over{2}} \left( {p_a^2 \over{a}} + a - \Lambda a^3 \right) \,.
\eqno(3.6)
$$
Variation with respect to $p_a$ recovers eq.(3.4), and variation with
respect to $N$ gives the constraint
$$
{\cal H} = 0 \,. \eqno(3.7)
$$
The corresponding equation of motion for $a$ is (for $N=1$)
$$
{\dot a}^2 +1-\Lambda a^2 =0 \,, \eqno(3.8)
$$
and its solution is the de Sitter space,
$$
a(t) = H^{-1} \cosh (Ht) \,, \eqno(3.9)
$$
where $H=\Lambda^{1/2}$.

Quantization of this model amounts to replacing $p_a \to -i\partial /\partial
a$ and imposing the Wheeler-De Witt equation
$$
\left[ {d^2 \over{da^2}} + {\gamma \over{a}}{d \over{da}} - U(a) \right]
\psi(a) = 0 \,, \eqno(3.10)
$$
where
$$
U(a) = a^2 (1-\Lambda a^2 ) \, \eqno(3.11)
$$
and the parameter $\gamma$ represents the ambiguity in the ordering of
non-commuting operators $a$ and $p_a$.  This equation is very similar to
eq.(2.13) for a nucleating bubble, and the following discussion closely
parallels that in Sec. 2.

The $\gamma$-dependent term in (3.10) does not affect the wave function in the
semiclassical regime.  Without this term, the equation has the form of a
one-dimensional Schrodinger equation for a "particle" described by a
coordinate $a(t)$, having zero energy and moving in a potential $U(a)$.
The classically allowed region is $a \ge H^{-1}$, and the WKB solutions of
eq.(3.10) in this region are
$$
\psi_{\pm} (a) = [p(a)]^{-1/2} \exp [\pm i\int_{H^{-1}}^a p(a')da' \mp i\pi/4 ]
\,, \eqno(3.12)
$$
where $p(a)=[-U(a)]^{1/2}$.  The under-barrier $a<H^{-1}$
solutions are
$$
{\tilde \psi}_{\pm} (a) = |p(a)|^{-1/2} \exp [\pm\int_a^{H^{-1}} |p(a')|da' ]
\,. \eqno(3.13)
$$
For $a>>H^{-1}$,
$$
{\hat p}_a \psi_{\pm} (a) \approx \pm p(a) \psi_{\pm} (a) \,, \eqno(3.14)
$$
and eq.(3.4) tells us that $\psi_- (a)$ and $\psi_= (a)$ describe an expanding
and a contracting universe, respectively
 (assuming that $N>0$).

In the tunneling picture, it is assumed that the universe originated at a small
size and then expanded to its
present, large size.  This means that the component of the wave function
describing a universe contracting from infinitely large size should be absent:
$$
\psi (a>H^{-1}) = \psi_- (a) \,. \eqno(3.15)
$$
The under-barrier wave function is found from the WKB connection formula,
$$
\psi(a<H^{-1}) = {\tilde \psi}_+ (a) - {i \over{2}}{\tilde \psi}_- (a) \,.
\eqno(3.16)
$$
Away from the
 classical turning point $a = H^{-1}$, the first term in (3.16) dominates, and
the nucleation probability can be approximated as\refto{linde, av84}
$$
\left| {\psi (H^{-1}) \over{\psi (0)}} \right|^2 \sim \exp \left(
-2\int_0^{H^{-1}} |p(a')| da' \right) = \exp \left( -{3 \over{8G^2 \rho_v}}
\right) \,. \eqno(3.17)
$$

It should be noted that the choice of $N>0$ in (3.4) is a matter of
convenience.  With the opposite choice, the roles of
$\psi_+ (a)$ and $\psi_-(a)$ would be reversed, and the boundary condition
(3.15) would be replaced by $\psi (a>H^{-1})=\psi_+(a)$.  This would result
in a time reversal
transformation $\psi(a) \to \psi^* (a)$.  Another way to look at this is to
note that the time coordinate $t$ is an arbitrary label in general relativity,
and it is a matter of convention to choose time growing or decreasing towards
the future (where
"future" is defined, e.g., by the growth of entropy or by the expansion of the
universe).  Clearly, there is no physical
ambiguity here, and once the convention is set, the tunneling wave function in
this model is uniquely defined.

At this point, I would like to mention the "generic" boundary condition
suggested by Strominger\refto{strom}.  He argued that since the nucleation
of the universe is governed by small-scale physics, the boundary condition on
$\psi$ should be imposed at small $a$, rather than at
large $a$ as in the tunneling approach.  The large-scale behavior of $\psi$
can then be determined without specifying the precise form of
this boundary condition.  The under-barrier wave function is generally given
by a linear combination of ${\tilde \psi}_+(a)$ and ${\tilde \psi}_-(a)$, and
for a
"generic" boundary condition at $a=0$, one expects the two terms to be
comparable at small $a$.  However, ${\tilde \psi}_+(a)$ decreases exponentially
 with $a$, while ${\tilde \psi}_-(a)$ exponentially grows and therefore
dominates for all but very small $a$.  The corresponding wave function in the
classically allowed range is found with the aid of the WKB connection formula:
$$
\psi (a<H^{-1}) = {\tilde \psi}_-(a) \,,
$$
$$
\psi (a>H^{-1}) = \psi_+(a) + \psi_-(a) \,. \eqno(3.18)
$$
The same wave function is obtained\refto{haw84} by applying
 the Hartle-Hawking prescription to this model.\refto{foot2}

The case for imposing boundary conditions at small $a$ appears to me
unconvincing.  The same argument
 could be applied to bubble nucleation, but there we know that the correct
boundary condition is the outgoing wave at large radii.  Another familiar case
when physics is confined to small scales while the boundary conditions are
imposed at infinity is a bound state, like the hydrogen atom.  In the next
section, we shall discuss how the tunneling wave function can be obtained by
analytic continuation from a "bound-state" universe.

\head{4. Tunneling Wave Function by Analytic Continuation}

The quantum-mechanical wave function for the decay of a metastable state is
often obtained by analytically continuing the bound state
wave function from the parameter values for which the corresponding state is
stable.  A similar approach can be adopted in quantum
cosmology.  As an example, we again consider the minisuperspace model (3.1),
but now with $\rho_v <0$.  In this case $\Lambda <0$, and it is clear
that the classical equation of motion (3.8) has no solutions.  However,
microscopic, Planck-size
universes could still pop out and collapse as quantum fluctuations.  Then one
expects the wave function to be peaked at very
small scales and to vanish at $a \to \infty$.

When dealing with analytic continuation, approximate solutions like (3.12),
(3.13) are
not sufficient, since the neglected terms can become large after continuation.
 We shall, therefore, use the exact solutions to
eq.(3.10) which can be obtained\refto{av85} for a particular choice of the
factor-ordering parameter, $\gamma =-1$.  With the boundary condition
$$
\psi (a \to \infty) =0 \,, \eqno(4.1)
$$
the solution is the Airy function
$$
\psi (a) = Ai (z) \,, \eqno(4.2)
$$
where
$$
z=(-2\Lambda)^{-2/3} (1-\Lambda a^2) \,. \eqno(4.3)
$$
The asymptotic behavior of (4.2) at large $a$ is
$$
\psi (a) \propto a^{-1/2} \exp [-(-\Lambda)^{1/2} a^3/3] \,. \eqno(4.4)
$$

Continuation to positive values of $\Lambda$ amounts to changing
$(-2\Lambda)^{-2/3} \to (2\Lambda)^{-2/3}\exp (\mp 2\pi i/3)$, where the sign
depends on the direction of rotation in the
complex $\Lambda$-plane.
Choosing the upper sign and using the relation\refto{absteg}
$$
2e^{\pm \pi i/3} {\rm Ai}(ze^{\mp 2 \pi i/3} ) = {\rm Ai}(z) \pm i {\rm Bi}
(z) \,,
\eqno(4.5)
$$
we conclude that the wave function for $\Lambda >0$ is
$$
\psi (a) = {\rm Ai}({\tilde z}) + i{\rm Bi}({\tilde z}) \,, \eqno(4.6)
$$
with
$$
{\tilde z} = (2\Lambda)^{-2/3} (1-\Lambda a^2) \,. \eqno(4.7)
$$
This is the tunneling wave function\refto{av88}.  The corresponding asymptotic
form at
 large $a$ is
$$
\psi (a) \propto a^{-1/2} \exp (-i\Lambda^{1/2} a^3/3 ) \,. \eqno(4.8)
$$

At this point, I would like to comment on one important difference between the
above analysis and the standard treatment of the decay of a metastable state.
In the standard approach, the Schrodinger equation for the bound state
of a particle,
$$
{\cal H}\psi = E\psi \,, \eqno(4.9)
$$
is solved with the boundary conditions $\psi \to 0$ at both $x \to \infty$ and
$x \to -\infty$.
The energy eigenvalues $E_n$ are then completely determined by the Hamiltonian
${\cal H} =-\partial_x^2 + U(x)$.  In the course of analytic continuation, as
the parameters of the potential $U(x)$ are changed, the eigenvalues $E_n$ also
change and develop imaginary parts as the corresponding
 states become metastable.  The resulting wave functions describe a probability
 that is exponentially decreasing with time inside the potential well by
gradually leaking to infinity. On the other hand, in the quantum-cosmological
model (3.10) the eigenvalue of the Wheeler-De Witt operator is fixed at E=0.
At the same time, the wave function is defined on a half-line $a>0$, and the
boundary condition (4.1) is imposed only at $a \to \infty$.  The wave function
is time-independent, and a steady probability flux at $a \to \infty$ is
sustained by an incoming flux through the boundary at $a=0$.  In fact, as
eq.(1.5) suggests, the tunneling wave function
is more appropriately thought of as a Green's function with a source at $a=0$,
rather than an eigenstate of the Wheeler-DeWitt
operator.  This will be further discussed in Sec.5,7.

We note finally that a "generic" choice of boundary condition
at $a=0$ would lead, for $\Lambda <0$, to a wave function which is not confined
 to small scales, but instead increases without bound at $a \to \infty$.

\head{5. Tunneling Wave Function from a Path Integral}

To discuss the relation between
 the outgoing-wave and path-integral forms of the tunneling proposal, we shall
consider a slightly more complicated minisuperspace model: a Robertson-Walker
universe with a homogeneous scalar field.  After appropriate rescalings of the
scalar field $\phi$ and
scale factor a, the corresponding Lagrangian and Hamiltonian can be written as
$$
{\cal L}={1\over{2}}[e^\alpha +e^{3\alpha}(-{\dot \alpha}^2 +{\dot \phi}^2
-V(\phi)] \,, \eqno(5.1)
$$
$$
{\cal H}={1\over{2}}[e^{-3\alpha}(-p_\alpha^2 +p_\phi^2 ) -e^\alpha
+e^{3\alpha}V(\phi)] \,. \eqno(5.2)
$$

Here, $\alpha = \ln a$, $V(\phi)$ is the scalar field potential,
 and the lapse function has been set $N=1$.

The path integral (1.4) for this model can be expressed in the
form\refto{teitel82, hallou89}
$$
K(q_2,q_1)=\int_0^\infty dT k(q_2,q_1;T) \,, \eqno(5.3)
$$
$$
k(q_2,q_1;T)=\int_{q_1}^{q_2} [dq]\exp \left( i\int_0^T {\cal L}dt \right)
\,,\eqno(5.4)
$$
where $q=(\alpha,\phi)$ and the integration is taken over all paths $\alpha
(t), \phi (t)$ beginning at $q_1 =(\alpha_1,\phi_1)$ at $t=0$ and ending at
$q_2 =(\alpha_2,\phi_2)$ at $t=T$.  The function $k(q_2,q_1,T)$
in eq. (5.4) has the familiar form of an amplitude for a "particle" to
propagate from $q_1$ to $q_2$ in time $T$ and satisfies the
Schrodinger equation
$$
\left(i{\partial \over{\partial T}} -{\cal H}_2 \right)k(q_2,q_1;T) =0 \,
\eqno(5.5)
$$
with the initial condition
$$
k(q_2,q_1;0) = \delta (q_2,q_1) \,. \eqno(5.6)
$$
The equation for $K(q_2,q_1)$ follows from (5.5), (5.6):
$$
{\cal H}_2 K(q_2,q_1) = -i\delta (q_2,q_1) \,. \eqno(5.7)
$$
Here, ${\cal H}$ is the Wheeler-DeWitt operator,
$$
{\cal H}={1\over{2}}e^{-3\alpha}[\partial_\alpha^2 - \partial_\phi^2
-U(\alpha,\phi)] \, \eqno(5.8)
$$
with "superpotential"
$$
U(\alpha,\phi)=e^{4\alpha}[1-e^{2\alpha}V(\phi)] \,, \eqno(5.9)
$$
and I am ignoring the factor-ordering ambiguity.  The subscript "2" of ${\cal
H}$ in (5.5) and (5.7) indicates that $\alpha$ and $\phi$ in (5.8) are taken to
be $\alpha_2$ and $\phi_2$.

Apart from an overall factor,
 the operator ${\cal H}$ in (5.8) is just the Klein-Gordon operator for a
relativistic "particle" in a (1+1)-dimensional "spacetime",
with $\phi$ playing the role of a spatial coordinate and $\alpha$ the role of
time.  The "particle" moves in an external potential $U(\alpha,\phi)$.  Let
us now consider the behavior of $K(q_2,q_1)$ as $\alpha_2 \to \pm \infty$ with
$\alpha_1$ fixed.  We must first note that for $\alpha \to -\infty$ the
potentital (5.9) vanishes, and $K(q_2,q_1)$ should be given\refto{foot3}
by a superposition of plane waves, $\exp [ik(\alpha_2 \pm \phi_2 )]$.
Since the path integral in (5.4) is taken over paths originating at some finite
$(\alpha_1,\phi_1)$ and going off to large negative $\alpha_2$, this
superposition should include only waves with $k>0$.  (Recall that $p_\alpha >0$
corresponds to ${\dot \alpha} <0$).

As $\alpha_2 \to +\infty$, the potential $U(\alpha,\phi)$ diverges, and the WKB
approximation becomes
increasingly accurate.  The dependence of $K(q_2,q_1)$ on $q_2$ is then given
by a superposition of terms $e^{iS}$, where $S$ is a solution of the
Hamilton-Jacobi equation,
$$
\left( {\partial S \over{\partial \alpha}}\right)^2 - \left( {\partial S
\over{\partial \phi}}\right)^2 +U(\alpha,\phi) =0 \,. \eqno(5.10)
$$
In each term, the function $S(\alpha,\phi)$ describes a congruence of classical
 paths with
$$
{d\phi \over{d\alpha}}=-{(\partial S/\partial\phi) \over{(\partial S/\partial
\alpha)}} \,. \eqno(5.11)
$$

For $V(\phi)>0$, $U(\alpha,\phi) \approx -e^{6\alpha}V(\phi) <0$, and it
follows from (5.11) that $|d\phi/d\alpha|<1$.  Hence, the "particle"
trajectories are asymptotically "timelike"
and correspond either to expanding universes with $p_\alpha =\partial
S/\partial \alpha <0$ or to universes contracting from an infinite size with
$\partial S/\partial \alpha >0$.  Since
all paths
originate at $\alpha_1 <\infty$, the superposition should include only terms
with $\partial S/\partial \alpha <0$.  For $V(\phi)<0$, the trajectories are
asymptotically "spacelike"
and cannot extend to timelike infinity ${i}_+$ or to null infinity ${\cal
I}_+$.  One expects, therefore, that $K(q_2,q_1) \to 0$ for $q_2$ at
${i}_+$ or ${\cal I}_+$.

Thus we see that the propagator $K(q_2,q_1)$ satisfies the outgoing-wave
boundary conditions both at $\alpha \to -\infty$ and $\alpha \to +\infty$.  The
 tunneling wave function (1.4)
is obtained by letting $\alpha_1 \to -\infty$ and integrating over all
initial values of $\phi$,
$$
\psi (\alpha,\phi) = \int_{-\infty}^{\infty} d\phi' K(\alpha,\phi |
-\infty,\phi') \,. \eqno(5.12)
$$
The trajectories then originate at the past timelike
infinity ${i}_-$, but the behavior of $\psi$ on the rest of the superspace
 boundary should be the same as that of $K$.  This is illustrated
 in Fig. 1 for the case of $V(\phi)>0$.  The probability flux is injected into
superspace at ${i}_-$ and exits in the form of outgoing waves
through ${\cal I}_-$ and ${i}_+$.  We conclude that the path-integral and
the outgoing-wave forms of the tunneling wave function are equivalent,
 at least in the simple model (5.1).  This is not very surprising, since
eqs.(5.3)-(5.7) coincide with the standard equations
for Feynman propagator in the proper-time representation, and the causal
boundary conditions for the propagator are the same as
the outgoing-wave boundary conditions for $\psi$.

\head{6. Beyond Minisuperspace}

The main difficulty in formulating the outgoing-wave boundary condition in the
general case is similar to the difficulty with the definition of
positive-frequency modes
 in a general curved spacetime.  There is, however, a hopeful sign.  Our
definition of outgoing waves in the minisuperspace model (5.1) was based on
rather general properties of the potential $U(\alpha,\phi)$:  its unbounded
growth at $\alpha \to +\infty$ and its vanishing at $\alpha \to -\infty$.  It
is
not difficult to verify that the superpotential in the Wheeler-DeWitt equation
has similar properties in the general case.

The general form of the Wheeler-DeWitt equation can be written as\refto{dewitt}
$$
(\nabla^2 -U)\psi =0 \,, \eqno(6.1)
$$
where
$$
\nabla^2 =\int d^3x N \left[ G_{ijkl}{\delta \over{\delta h_{ij}}} {\delta
\over{\delta h_{kl}}} +{1\over{2}}h^{-1/2}{\delta^2 \over{\delta \phi^2}}
\right] \, \eqno(6.2)
$$
is the superspace Laplacian,
$$
G_{ijkl} = {1 \over{2}}h^{-1/2}(h_{ik}h_{jl} +h_{il}h_{jk} -h_{ij}h_{kl}) \,
\eqno(6.3)
$$
is the superspace metric,
$$
U=\int d^3xNh^{1/2} [- R^{(3)} + {1\over{2}}h^{ij}\phi_{,i}\phi_{,j} +V(\phi)]
\, \eqno(6.4)
$$
is the superpotential, $h_{ij}(x)$ and $N(x)$ are, respectively, the 3-metric
and the lapse function in the (3+1)
decomposition of spacetime,
$$
ds^2=(N^2 +N_iN^i)dt^2 -2N_idx^idt-h_{ij}dx^idx^j \,, \eqno(6.5)
$$
$h=\det (h_{ij})$ and $R^{(3)}$ is the curvature of 3-space.  As before,
matter fields are represented by a single scalar
field $\phi$ and I have ignored the factor-ordering  problem.  The wave
function $\psi$ is a function of $h_{ij}(x)$ and $\phi(x)$, but is independent
of $N(x)$. The metric
$h_{ij}$ can be represented as
$$
h_{ij}=e^{2\alpha}{\tilde h}_{ij} \,, \eqno(6.6)
$$
where $\det ({\tilde h}_{ij})=1$.  Then the Laplacian term in (6.1) is $\propto
\exp (-3\alpha)$, the first two terms in the superpotential (6.4) are $\propto
\exp (\alpha)$, and the last term is
$\propto \exp (3\alpha)$.  The relative magnitude of these terms for $\alpha
\to \pm \infty$ is the same as
 in eq. (5.9).

In the limit $\alpha \to -\infty$, the superpotential $U$ in (6.1) becomes
 negligible, and one can hope to define outgoing modes analogous to the plane
waves of the previous section.  This
possibility has also been suggested by Wald\refto{wald} in a different context.
 Here,
 I will not attempt to analyze the most general case
and illustrate the idea in a reduced superspace model which includes all
degrees of freedom of the scalar field $\phi$, but only
one gravitational variable $\alpha$.  With the scalar field represented as
$$
\phi (x)=(2\pi^2)^{1/2} \sum_n f_nQ_n (x) \,, \eqno(6.7)
$$
where $Q_n(x)$ are the harmonics on a 3-sphere, the superspace
Laplacian
(6.2) takes the form\refto{halhaw, vv88}
$$
\nabla^2 = e^{-3\alpha}\left( {\partial^2 \over{\partial \alpha^2}} - \sum_n
{\partial^2 \over{\partial f_n^2}} \right) \,. \eqno(6.8)
$$
The plane-wave asymptotic solutions are then
$$
\psi (\alpha, f_n) =\exp (ik_\alpha \alpha +i\sum_n k_n f_n) \, \eqno(6.9)
$$
with
$$
k_\alpha^2 -\sum_n k_n^2 =0 \,. \eqno(6.10)
$$
The tunneling wave function includes only terms with $k_\alpha >0$.  This is
the boundary condition at $\alpha \to -\infty$.

To formulate the tunneling condition on the remainder of superspace boundary,
one first has to specify what that boundary is.  In other words, we should
decide what class of metrics and
matter fields should be included in superspace.  The form of the Wheeler-DeWitt
 equation (6.1)-(6.4) suggests that we should
include all configurations $\{h_{ij}(x),\phi (x)\}$ for which $h^{1/2}R^{(3)},
h^{1/2}h^{ij}\phi_{,i}\phi_{,j}$ and $h^{1/2}V(\phi)$ are integrable functions.
 Then the superpotential $U$ is finite
everywhere in
superspace and will generically diverge towards the boundary.  This happens,
in particular, at $\alpha \to +\infty$.
As $|U| \to \infty$, some components of the
gradient $\nabla S$ in the Hamilton-Jacobi equation
$$
(\nabla S)^2 +U=0 \, \eqno(6.11)
$$
should also diverge, and one can hope that the WKB approximation will become
asymptotically exact, thus allowing one to define outgoing waves.\refto{foot4,
foot5}  For example, when some dimensions of the universe become very
large (e.g., $\alpha \to \infty$), the classical description of the
corresponding degrees of freedom becomes increasingly accurate.  Denoting these
 classical variables by $c_i$; and the remaining variables
 by $q_j$, the asymptotic form of the wave function can be written
as\refto{halhaw, laprub, banks85, av89}
$$
\psi (c,q) =\sum_N e^{iS_N (c)} \chi_N (c,q) \,. \eqno(6.12)
$$
The Hamilton-Jacobi functions $S_N(c)$ describe
congruences of classical paths, $p_i =-\partial S/\partial c_i$.  The tunneling
 boundary condition selects the solutions of (6.10) which include only outgoing
 paths,
evolving towards the boundary.

It should be noted that superspace defined by the
condition $|U|<\infty$ includes a very wide class of configurations.  The
metric and matter fields have to be continuous, but not necessarily
differentiable.  In particular, scalar fields with discontinuous derivatives
and metrics with $\delta$-function curvature
singularities on surfaces, lines, and points are acceptable configurations.
This conclusion fits well with the path integral approach,
where it is known that the path integral is dominated by the paths which are
continuous, but not differentiable.  The superspace
configurations can be thought of as slices of these paths.

If the definition of outgoing waves along the lines indicated
in this section is indeed possible, then the same argument as in Section 5
suggests that the wave function defined by the path
integral (1.4) should satisfy the outgoing-wave boundary condition.  One
advantage of the path-integral definition is that it
may be consistent even if outgoing waves cannot always be defined.  Another
advantage is that the path integral version appears to
 be better suited to handle topology change (see Sec. 7).

\head{7. Topology change}

In the discussion, so far, I have not touched upon the issue of topology change
in quantum gravity.  This issue, however, can hardly  be avoided, since the
"creation of a universe from nothing" is an example of topology-changing event.

The Wheeler-DeWitt equation (6.1) is based on canonical quantum gravity, which
assumes the spacetime to be a manifold of topology $R \times \Sigma$, where
$R$ is a real line and $\Sigma$ is a closed 3-manifold of arbitrary but fixed
topology.  The corresponding superspace ${\cal G}_\Sigma$ includes only
3-metrics of topology $\Sigma$.  We can define the extended superspace ${\cal
G}$ including all possible topologies.  It can be split into topological
sectors, with all metrics in each sector having the same topology.

The division of superspace into topological
 sectors can be illustrated by lower-dimensional examples.  In the
(1+1)-dimensional case, 3-geometries are replaced by lines (strings), and
topological sectors can simply be labeled by the occupation number of closed
strings.  In (2+1) dimensions, a point
 $g \in {\cal G}$ corresponds to a number of closed surfaces (membranes), and
each surface can be characterized by the number of handles.\refto{genus}
Each
topological sector of ${\cal G}$ can thus be labeled by an infinite set of
integers $\{n_0, n_1, ... \}$ giving respectively the occupation
numbers for surfaces with  $0,1, ...$ handles.  In the (3+1)-dimensional case,
there is a much richer structure, but a topological classification of
3-dimensional manifolds has not yet been given.

Creation of a universe from nothing described in Sections 3-5 is a transition
from the null topological sector containing no universes at all to the sector
with one universe of topology $S_3$.  The surface $\alpha =-\infty, |\phi|
<\infty$ can be thought of as a boundary between the two sectors.  The
probability flux is injected into superspace through this boundary (see Fig.1)
and flows out of superspace through the remaining boundary ($\alpha \to
-\infty$
with $|\phi| \to \infty$, or $\alpha \to +\infty$).  One could have thought
that topology-changing transitions always occur through
the boundaries of the corresponding superspace sectors.  This was the point of
view I adopted in my earlier formulation of the tunneling boundary
condition\refto{av88}. I no longer believe this picture to be correct, but it
may still be useful in cases when topology change is a semiclassical tunneling
event.  In this section I shall first review the motivation for the old
approach, then explain why I think it is not applicable in the general case,
and finally discuss some alternative approaches to topology change.

Tunneling amplitudes in quantum field theory are often evaluated
semiclassically using the steepest descent approximation.  One then finds
that the path integral for the amplitude is dominated by a solution of
Euclidean field equations, called the instanton.  If topology change is a
quantum tunneling event, one can similarly expect it to be represented by a
smooth Euclidean manifold, ${\cal M}$, interpolating between the initial
configuration $\Sigma_1$ and the final configuration, $\Sigma_2$.  The
intermediate superspace configurations can be obtained as slices of ${\cal M}$
and can be conveniently described using the concepts of Morse
theory\refto{milnor}.

Consider a smooth real function $f(x)$ on manifold ${\cal M}$.  A point $x_0$
is called a critical
 point of $f$ if $\partial_\mu f(x_0) =0$.  A critical point is called
non-degenerate if $det [\partial_\mu \partial_\nu f(x_0)] \not= 0$.  We shall
call $f(x)$ a Morse function if it has the following properties:  (i) $f(x)$
takes values between $0$ and $1$, with  $f(x)=0 ~~iff~~x \in \Sigma_1$, and
$f(x)=1~~iff~~x \in \Sigma_2$; (ii) all critical points of $f$ are in the
interior
of ${\cal M}$ (that is, not on the boundary) and are non-degenerate.  In a 2-d
example of Fig. 2, the manifold ${\cal M}$ is shown embedded in a
 3-dimensional space, and the Morse function is given by the projection on the
vertical axis.  In this case, the saddle point $P$ is a critical point of
$f(x)$.
It can be shown that a Morse function can always be defined and that it always
has some critical points if $\Sigma_1$ and $\Sigma_2$ have different topology.
 We shall assume that $f(x)$ is chosen so that it has the smallest possible
number of critical points, that is, no more than dictated by topology.

Slices of ${\cal M}$ corresponding to superspace configurations can be
obtained as surfaces of constant $f$.  (Different choices of Morse function
will, of course, give different slicings).
 These slices will have a smooth geometry, except the critical slices
passing through critical points.  With an appropriate choice of
locally-Cartesian coordinates, the Morse function in the vicinity of the
critical point
can be represented as
$$
f(x) = \sum_{i=1}^d a_i x_i^2 \,, \eqno(7.1)
$$
where $d$ is the dimensionality of space ($d = 3$).  The
critical section, $f(x) = 0$, is a generalized cone. For $d \ge 3$ it has a
curvature singularity of the form\refto{sergio}
$$
R \propto r^{-2} \,, \eqno(7.2)
$$
where $r$ is the distance from the critical point, $r^2 = \sum_i x_i^2$.  For
$d = 2$, the curvature has a $\delta$-function singularity, $R^{(2)} h^{1/2}
\propto \delta^{(2)} (x)$.
An important special case
of topology change is the "creation of universes from nothing", when the
initial configuration is absent.  A 2-dimensional illustration is shown in
Fig. 3.
Here, the critical slice is a single point.  For near-critical slices, in $d
\ge 2$ the curvature is again given by (7.2), where now $r$ is the
characteristic size of $d$-space.

The idea of Ref.\cite{av88} was
that the boundary of superspace can be divided into regular and singular parts.
  The regular boundary includes only configurations which can be obtained as
critical slices of smooth Euclidean manifolds.  Such configurations correspond
to transitions between different topological sectors.  The remaining part of
the boundary is called the singular boundary, and the outgoing-wave boundary
condition is imposed only on that part.  The boundary condition on regular
boundary was supposed to enforce conservation of probability flux as it flows
from one topological sector to another, but no specific form of the boundary
condition was proposed.  The overall picture was that
the probability flux is injected into superspace through the boundary with the
null sector, it then flows between different topological sectors through the
regular boundaries, and finally flows out of superspace through the singular
boundary.\refto{foot6}

As I mentioned earlier, I no longer think this picture can be valid in the
general case.  The main reason is that topology change does not necessarily
occur between configurations at the boundaries of superspace sectors, but
generally involves
configurations in the interior of these sectors.  It is true that, in order to
change topology, one has to go through a singular
 3-geometry.  But, as we discussed in Section 6, superspace includes a very
wide class of configurations, such as metrics with integrable curvature
singularities and scalar fields with discontinuous derivatives.  Note in
particular that curvature singularities (7.2) on critical slices are
integrable, and therefore the critical slices will generally lie in the
interior of superspace.

To give a specific example, consider creation of a wormhole in a universe
having initially the topology of $S_3$.  The
transition is then between the topological sectors $S_3$ and $S_1 \times S_2$.
 The wormhole radius can be defined as $r = (A_{min}/4\pi)^{1/2}$, where
$A_{min}$ is the smallest cross-sectional area of the wormhole, and can be
used as one of superspace variables.  Since $r$ has a semi-infinite range,
$r=0$ is a superspace boundary in the sector $S_1 \times S_2$.  On dimensional
grounds, the curvature in the wormhole vicinity is $R^{(3)} \sim r^{-2}$, and
the integral of $R^{(3)} h^{1/2}$ does not diverge as $r \to 0$.  The boundary
at $r=0$ is therefore similar to what was called the regular boundary in
Ref.\cite{av88}.  On the other hand, configurations in the $S_3$ sector
"right before" topology change do not lie on any boundary. These configurations
should only satisfy the continuity requirement: all matter fields should take
the same values at the points that are about to be identified.

An important example of topology change in lower dimensions is reconnection of
intersecting strings.  At the classical level, this process plays a crucial
role in the evolution of cosmic strings\refto{avbook} (see Fig.4).  At the
quantum level, it represents the elementary
interaction vertex in fundamental string theories.
A string loop can be thought of as a one-dimensional closed universe.  The
superspace configurations for the loop are given by
the functions $x^\mu (\sigma)$, where $\sigma$ is a parameter on the loop and
the spacetime coordinates $x^\mu$ play the role of worldsheet scalar fields.
Topology change (loop splitting) can occur in configurations where the loop
self-intersects, that is, when $x^\mu (\sigma_1) = x^\mu (\sigma_2)$ for some
$\sigma_1 , \sigma_2$.  These configurations are not special in any other way
and do not lie on superspace boundary.  The configurations immediately after
splitting have discontinuous derivatives of $x^\mu (\sigma)$ at reconnection
points.  They are also legitimate superspace configurations and do not belong
to a boundary.\refto{foot7}

The conclusion is that topology-changing transitions affect not only superspace
boundary, but can occur between points in the interior of different topological
 sectors.  This has an important implication:  in order to account for topology
 change, the Wheeler-DeWitt equation has to be modified.  In the tradition of
the subject, I would like to offer some speculations regarding the form of this
modified equation.

My suggestion is that the Wheeler-DeWitt operator ${\cal H}$ in (1.2) should be
modified by adding an operator ${\tilde \delta}$ that has matrix elements
between different superspace sectors.  The corresponding action can be written
symbolically as
$$
S = \int [dh] \psi^{*} {\cal H} \psi +
\int [dh][dh']\psi^{*} (h){\tilde \delta} (h,h')\psi (h') \,, \eqno(7.3)
$$
where the integration is taken over all superspace sectors and $h$ stands for
all
superspace variables.  It seems reasonable to assume that topology change is a
local process, then we should have ${\tilde \delta}(h_1 ,h_2)=0$ unless $h_1$
and $h_2$ can be obtained from
one another by changing topological relations at a single point.  The
Wheeler-DeWitt equation for $\psi_N (h)$ in topological sector $N$ is obtained
by varying (7.3):
$$
{\cal H}\psi_N (h) + \sum_{N' \not= N} \int [dh']{\tilde \delta}_{NN'}(h,h')
\psi_{N'}(h') = 0 \,. \eqno(7.4)
$$

The form of the operator ${\tilde \delta}(h,h')$ is, of course,
unknown.  One can hope to gain some insight into it by studying
lower-dimensional examples.  In the case of strings, the topological sectors
can be labelled by the number of disconnected loops, $n$, and ${\tilde
\delta}_{nn'}$ has matrix elements with $n'=n \pm 1$.  An even simpler example
is given by pointlike particles, which can be thought of as (0+1)-dimensional
universes with spacetime coordinates $x^\mu$ playing
the role of scalar fields and the Klein-Gordon operator $\nabla^2 + m^2$
playing the role of the Wheeler-DeWitt operator.  Topology change corresponds
to elementary particle
interactions, like the one illustrated in Fig.5 for a $\lambda \phi^3$
theory.  Here, two paticles merge into one, and the ${\tilde \delta}$ operator
should be proportional to $\delta$-functions ensuring that the initial and
final particles have the same coordinates at the moment of interaction.
It would be interesting to develop the first-quantized formalism for particle
interactions in the form (7.4) and verify its equivalence to a quantum field
theory with non-linear interactions.  The possibility of equivalence between a
linear system of equations (7.4) and a non-linear field theory may
seem rather unlikely.  It is well known, however, that the full content of a
perturbative quantum field theory can be expressed as an infinite set of
linear relations between the Green's functions (Schwinger-Dyson
equations).  A similar representation has also been obtained in matrix models
of two-dimensional quantum gravity\refto{susskind}.

To formulate the boundary conditions for the functions $\psi_N (h)$ in (7.4),
we again divide the superspace boundary into singular and regular parts.  The
singular boundary includes configurations with $|U| \to \infty$ and the null
part of the boundary at $\alpha \to -\infty$ (see Section 6).  The functions
$\psi_N$ should have only
outgoing waves at the singular boundary.  These waves carry the probability
flux,
$$
J_N = i(\psi_N^{*} \nabla \psi_N -\psi_N \nabla \psi_N^{*}) \,, \eqno(7.5)
$$
out of superspace.  The waves flowing into
and out of the regular boundary correspond to transitions between topological
sectors.  In the example of the $S_1 \times S_2 \to S_3$ transition, the
 flux flowing into the regular part of the boundary at $r=0$ in the $S_1 \times
S_2$ sector reappears through the source term on the right-hand
side of (7.4) in the $S_3$ sector.  The boundary condition at $r=0$ should
enforce flux conservation between the two sectors.  I will
 not attempt to write down a specific form of this boundary condition.

Assuming that outgoing waves can be defined along the lines of the previous
section and that the flux conservation condition is formulated, one can hope
that the wave function defined by eqs. (7.4) is equivalent to the one given by
the path integral (1.3), where the integration is performed over 4-manifolds
of arbitrary topology.  It is known that any Lorentzian metric interpolating
between two compact spacelike surfaces of different topology must either be
singular or contain closed timelike curves\refto{geroch}.  The singularities,
however,
can be very mild\refto{horow}, and there seems to be no reason for excluding
the
corresponding spacetimes from the path integral.  If all metrics of finite
 action are included, this would be more than sufficient to permit Lorentzian
topology change.

\head{8. Comments on Third Quantization}

It has often been argued\refto{rub88, mcgui, hosmor, gistro, banks88, fkps}
that an adequate description of topology change can be given in the third-
quantization approach, where the wave function $\psi$ is promoted to the status
of a quantum field operator.  Topology change is then accounted for by self-
interaction of $\psi$.
  For example, a $\psi^3$ interaction allows a parent universe, say of
topology $S_3$, to split into two daughter universes of the same topology.
This is probably adequate for one-dimensional universes
(strings), where topology is characterized simply by the
occupation number of closed loops.  However, in higher dimensions the situation
is not so simple.  For two-dimensional universes, one would have to introduce
an additional field creating and annihilating handles, while three-dimensional
topologies have not yet been classified, and one may need to introduce an
infinite number of fields and interaction types.  It is not evident, therefore,
that third quantization offers any advantages in describing topology change,
compared to the "first quantized" approaches like (7.4) or (1.4).

I would also like to comment on the specfic implementation of the third
quantization picture in simple minisuperspace models\refto{rub88, hosmor,
fkps}.  Without introducing non-linearity, the creation of universes in this
approach is described in a manner similar to the description of particle
creation in a time-varying external field.  The idea is suggested by the fact
that the Wheeler-DeWitt equation is similar to Klein-Gordon equation with the
scale variable $\alpha$ playing the role of time and the superpotential $U$
playing the role of a time-dependent external potential.  For $\alpha \to
-\infty$ the potential vanishes (see Sec. 6), and one can expand the filed
operator $\psi$ into positive and negative-frequency modes,
$$
\psi = \sum_k (a_k \psi_k + a_k^+ \psi_k^*)  \, \eqno(8.1)
$$
with $\psi_k (\alpha \to -\infty) \propto \exp (i\omega_k \alpha)$ , $\omega_k
>0$ and creation and annihilation operators satisfying the usual commutation
relations.  The
 "in-vacuum" state, containing no universes at $\alpha \to -\infty$, would
then be defined by $a_k |0>_{in} =0$, and single-universe states would be
given by $|k> = a_k^+ |0>$.
  In the opposite limit of $\alpha \to +\infty$, one can similarly define a
complete set of mode functions ${\tilde \psi}_k , {\tilde \psi}_k^*$, such
that ${\tilde \psi}_k (\alpha \to +\infty) \propto \exp (iS)$ with $\partial
S/\partial \alpha >0$, and write
$$
\psi = \sum_k ({\tilde a}_k {\tilde \psi}_k + {\tilde a}_k^+ {\tilde \psi}_k^*)
\,. \eqno(8.2)
$$
The state containing no universes at $\alpha \to +\infty$ is then $|0>_{out}$
with ${\tilde a}_k |0>_{out} =0$, and single-universe states are ${\tilde
a}_k^+ |0>_{out}$.

Since both sets of functions are complete, they must be linearly related to
one another,
$$
{\tilde \psi}_k = \sum_{k'} (\alpha_{kk'}\psi_{k'} + \beta_{kk'}\psi_{k'}^* )
\,, \eqno(8.3)
$$
and eqs. (8.1), (8.2) then imply a linear relation between the creation and
annihilaiton operators,
$$
{\tilde a}_k = \sum_{k'} (\alpha_{kk'}^* a_{k'} - \beta_{kk'}^* a_{k'}^+ ) \,.
\eqno(8.4)
$$
If the universal field $\psi$ is in the state $|0>_{in}$ containing no
universes at $\alpha \to -\infty$, then the
average number of universes in state $k$ at $\alpha \to +\infty$ is generally
non-zero and is given by
$$
<{\tilde n}_k > = <0|{\tilde a}_k^+ {\tilde a}_k |0>_{in} = \sum_{k'}
|\beta_{kk'} |^2 \,. \eqno(8.5)
$$
The suggestion in Refs.\refto{rub88, hosmor, fkps} is that $<{\tilde n}_k >$
should be interpreted as the number of universes created from nothing.  I
disagree with this interpretation for the reasons that I will now explain.

In the third quantization picture, there are no universes of vanishing size
$(\alpha \to -\infty)$, and as $\alpha$ grows, the number of universes
increases and finally reaches
its asymptotic value $<{\tilde n}_k >$ at $\alpha \to +\infty$.  The universes
are created at finite values of $\alpha$, that is, with a finite size.  This is
 drastically different from the creation-from-nothing picture, where the
universes start at zero size and continuously evolve towards larger sizes, so
that all the "creation" occurs at
 $\alpha \to -\infty$.

The origin of the difference between the two pictures is in the fact that the
"time" $\alpha$ is not really a monotonic
 variable: the universes can both expand and contract.  The positive- and
negative-frequency mode functions $\psi_k$ and $\psi_k^*$ correspond,
respectively, to expanding and contracting universes.  From this point of view,
 what is described in third quantization as creation of a pair of universes at
some $\alpha =\alpha_0$, is simply a contracting universe that turns around
and starts re-expanding at $\alpha =\alpha_0$.

This can be illustrated using a (0+1)-dimensional example: pair creation in an
external field.  Following Feynman, antiparticles can be interpreted as
particles travelling backwards in time, and pair creation corresponds to a
particle trajectory like the one
shown in Fig.6.  The trajectory can be represented as $x^\mu (\tau)$ with
$-\infty < \tau <\infty$.  Using the string theory language, $\tau$ is a
worldsheet time coordinate, and $x^\mu$ are target space coordinates.  For an
observer riding on the particle, $\tau$ is a suitable time coordinate and
$x^\mu (\tau)$ is a set of interacting scalar fields.\refto{hen}  The field
$x^0 (\tau)$ decreases with $\tau$ at $\tau \to -\infty$ and grows at $\tau \to
+\infty$.  On the other hand, an external (e.g., human) observer, whose home is
in the target space, will use $x^0$ as his time coordinate.

In the third quantization picture, the variable $\alpha$ plays the role of
target-space time, $x^0$.  It is not impossible that some super-human observer
living in this target space will observe the creation of pairs of
universes.\refto{espace}
However, we are interested in what happens
 from the point of view of a worldsheet observer, living inside the universe
and using the worldsheet time $\tau$. In any case, it appears that the process
described by the third quantization formalism (8.1-5) does not correspond to a
topology-changing nucleation of the universe that the authors of\refto{rub88,
hosmor, fkps} had in mind.

\head{9. Conclusions}

The wave function of the universe $\psi$ can be obtained either by solving the
Wheeler - De Witt equation with appropriate boundary conditions or by
performing a path integral over an appropriate class of paths.  Our discussion
in this paper was focussed on the tunneling proposal for $\psi$.  Although
little was proved, our discussion lead to several conjectures which will be
briefly summarized here.

In the path integral approach, the tunneling wave function is defined as a sum
over Lorentzian 4-geometries interpolating between a vanishing 3-geometry (a
point) and given 3-geometry.  The sum is, in general, performed over manifolds
of arbitrary topology.  I have argued that the wave function defined in this
way should satisfy the outgoing - wave condition on a part of superspace
boundary.

Superspace can be divided into topological sectors, and part of its boundary
can be thought of as the boundary between different sectors.  We call it
regular boundary. The rest of the boundary, which includes "incurably" singular
configurations, is called singular boundary (see Sec. 6 for more details).  The
outgoing - wave condition should be satisfied only on the singular boundary.  I
have argued that the superpotential (6.4) of the Wheeler - De Witt equation
either vanishes or diverges almost everywhere at this boundary and that this
may enable one to
give a precise definition of outgoing waves.

If the topology of the universe is restricted to be that of a sphere, then the
outgoing - wave boundary condition may be sufficient to determine the tunneling
wave function.  However, in the general case, this condition has to be
supplemented by some boundary conditions at the regular boundary.

If topology change is allowed, then I have argued that it will occur not only
through the boundaries between the superspace sectors, but will generally
involve configurations in the superspace interior.  This will result in a
modification of the Wheeler - De Witt equation.  A possible form of the
modified equation is suggested in Sec. 7.

Apart from the tunneling approach, I gave a critical discussion of the
"generic" boundary condition (in Sec. 3) and of the third quantization picture
(in Sec. 8).

The tunneling approach to the wave function of the universe was motivated by
the analogy with bubble nucleation and we may still gain important insights
into the complicated issues of quantum cosmology by studying the wave function
of the nucleating bubble.  We may also learn a great deal from quantum gravity
in two dimensions, which can be thought of as quantum cosmology of
one-dimensional closed universes (strings), and even from the ordinary quantum
field theory, in which the branching propagator lines in Feynman diagrams can
be thought of as branching 0-dimensional universes (particles).  However, in
pursuing these analogies, one should remember that in all these cases the
observer is usually assumed to be in the target space, while in quantum
cosmology the observer lives on the worldsheet.  The relation between the wave
functions of the universe (bubble, string, particle) obtained by these
different observers is an intriguing problem for future research.

\head{Acknowlegements}

I am grateful to Valery Rubakov and Andy Strominger for discussions and
comments
and to Alan Guth for his
hospitality at MIT.  I also acknowlege support by the National Science
Foundation and by the U.S. Department of Energy.

\references

\refis{dewitt} B. S. De Witt, Phys. Rev. {\bf 160}, 1113 (1967)

\refis{tryon} E. P. Tryon, Nature {\bf 246}, 396 (1973)

\refis{fomin} P. I. Fomin, Dokl. Akad. Nauk Ukr. SSR {\bf 9A}, 831 (1975)

\refis{atpag} D. Atkatz and H. Pagels, Phys. Rev. {\bf D25}, 2065 (1982)

\refis{av82} A. Vilenkin, Phys. Lett. {\bf 117B}, 25 (1982)

\refis{grizel} L. P. Grishchuk and Ya. B. Zel'dovich, in " Quantum Structure of
Space and Time", ed. by M. Duff and C. Isham, Cambridge University Press,
Cambridge (1982)

\refis{hh} J. B. Hartle and S. W. Hawking, Phys. Rev. {\bf D28}, 2960 (1983)

\refis{linde} A. D. Linde, Lett. Nuovo Cim. {\bf 39}, 401 (1984)

\refis{av84} A. Vilenkin, Phys. Rev. {\bf D30}, 509 (1984)

\refis{ghp} G. W. Gibbons, S. W. Hawking and M. J. Perry, Nucl. Phys. {\bf
B138}, 141 (1978)

\refis{teitel80} C. Teitelboim, Phys. Lett. {\bf B96}, 77 (1980)

\refis{teitel82} C. Teitelboim, Phys. Rev. {\bf D25}, 3159 (1982)

\refis{halhar} J. J. Halliwell and J. B. Hartle, Phys. Rev. {\bf D41}, 1815
(1990)

\refis{av86} A. Vilenkin, Phys. Rev. {\bf D33}, 3560 (1986)

\refis{av88} A. Vilenkin, Phys. Rev. {\bf D37}, 888 (1988)

\refis{hallou89} J. J. Halliwell and J. Louko, Phys. Rev. {\bf D39}, 2206
(1989)

\refis{rub88} V. A. Rubakov, Phys. Lett. {\bf 214B}, 503 (1988)

\refis{rub84} V. A. Rubakov, Phys. Lett. {\bf 148B}, 280 (1984)

\refis{banks88} T. Banks, Nucl. Phys. {\bf B309}, 493 (1988)

\refis{gistro} S. B. Giddings and A. Strominger, Nucl. Phys. {\bf B321}, 481
(1989)

\refis{mcgui} M. McGuigan, Phys. Rev. {\bf D38}, 3031 (1988)

\refis{hosmor} A. Hosoya and M. Morikawa, Phys. Rev. {\bf D39}, 1123 (1989)

\refis{fkps} W. Fischler, I. Klebanov, J. Polchinski and L. Susskind, Nucl.
Phys. {\bf B327}, 157 (1989)

\refis{llqm} L. D. Landau and E. M. Lifshitz, Quantum Mechanics, (Pergamon,
Oxford, 1965)

\refis{vko} M. B. Voloshin, I. Yu. Kobzarev and L. B. Okun, Yad. Fiz. {\bf 20},
1229 (1974) [Sov. J. Nucl. Phys. {\bf 20}, 644 (1975)]

\refis{col} S. Coleman, Phys. Rev. {\bf D15}, 2929 (1977)

\refis{calcol} C. G. Callan and S. Coleman, Phys. Rev. {\bf D16}, 1662 (1977)

\refis{vv91} T. Vachaspati and A. Vilenkin, Phys. Rev. {\bf D43}, 3846 (1991)

\refis{gv92} J. Garriga and A. Vilenkin, Phys. Rev. {\bf D45}, 3469 (1992)

\refis{halhaw} J. J. Halliwell and S. W. Hawking, Phys. Rev. {\bf D31}, 1777
(1985)

\refis{vv88} T. Vachaspati and A. Vilenkin, Phys. Rev. {\bf D37}, 898 (1988)

\refis{foot1} Although there seems to be no dispute about this result, the
exact form of the  boundary condition that selects this wave function is not
clear.  The non-perturbative, radial part of $\psi$ should certainly satisfy
the outgoing-wave boundary condition, and it was argued in Refs.\refto{vv91,
gv92} that the rest of $\psi$ should be fixed by the regularity condition,
$|\psi| < \infty$.  However, it was later realized\refto{vach} that this is not
sufficient:  the wave function obtained from $\psi$ by acting with
$\Phi$-particle creation operators still satisfies all the boundary conditions.
Sasaki {\it et.al.}\refto{tsy, styy} emphasized that the boundary condition
should reflect the fact that the bubble nucleates from vacuum and not from some
other excited state.  However, the specific form of the
boundary condition they suggest is not suitable for a thin-wall bubble.  It is
possible that $\psi$ can be completely fixed only by
requiring that it respects the Lorentz invariance of the false vacuum.

 \refis{vach} T. Vachaspati, unpublished

\refis{tsy} T. Tanaka, M. Sasaki and K. Yamamoto, Phys. Rev. D, in press

\refis{styy} M. Sasaki, T. Tanaka, K. Yamamoto, and J. Yokoyama, Prog. Theor.
Phys., in press

\refis{strom} A. Strominger, Nucl. Phys. {\bf B321}, 481 (1989)

\refis{haw84} S. W. Hawking, Nucl. Phys. {\bf B239}, 257 (1984)

\refis{foot2} The wave function (3.18) includes expanding and contracting
components with equal amplitudes, and it is natural to interpret it as
describing a contracting and re-expanding de Sitter universe (3.9).  An
alternative view\refto{page, rub88}, is to interpret $\psi_-(a)$ and its time-
reverse $\psi_+(a)$ as describing the same nucleating universe, but with a
different choice made for the direction of the time coordinate.  This
interpretation may be problematic due to interference between the two
components of $\psi$.

\refis{page} D. N. Page, Phys. Rev. {\bf D32}, 2496 (1985)

\refis{av85} A. Vilenkin, Nucl. Phys. {\bf B252}, 141 (1985)

\refis{absteg} M. Abramovitz and I. A. Stegun, Handbook of Mathematical
Functions.

\refis{susskind} For a discussion in the context of two-dimensional quantum
cosmology, see A.Cooper, L. Susskind and L. Thorlacius, Nucl. Phys. {\bf B363},
132 (1991)

\refis{foot3} I assume that $V(\phi)$ grows slower than $\exp (6|\phi|)$ at
$\phi \to \pm \infty $.

\refis{wald} R. Wald, Phys. Rev. {\bf D37}, 888 (1988)

\refis{foot4} Whether or not this actually happens, depends  on how fast $U$
diverges towards the boundary.  The WKB approximation  assumes that $|\nabla^2
S |<<|(\nabla S)^2 |$.  With $\nabla S \sim U^{1/2}$, this implies $|\nabla U|
<<|U|^{3/2}$.

\refis{foot5} We note in passing, that the wave function (6.9) in the $\alpha
\to -\infty$ limit is also of the semiclassical form $\exp (iS)$ and that the
WKB approximation becomes increasingly accurate as $\alpha \to -\infty$.

\refis{laprub} V. Lapchinsky and V. A. Rubakov, Acta Phys. Pol. {\bf B10}, 1041
(1979)

\refis{banks85} T. Banks, Nucl. Phys. {\bf B249}, 332 (1985)

\refis{av89} A. Vilenkin, Phys. Rev. {\bf D39}, 1116 (1989)

\refis{milnor} J. Milnor, "Lectures on h-Cobordism Theorem", Princeton
University Press (1965)

\refis{sergio} S. del Campo and A. Vilenkin, unpublished

\refis{foot6} In Ref.\cite{av88} I defined the regular boundary as consisting
 of singular configurations which can be obtained by slicing regular Euclidean
4-geometries, but the relation to Morse functions was not spelled out.  This is
somewhat imprecise and has lead to mis-interpretations.\refto{louvach,
hallou90, ugl}
Consider, for example, a manifold
${\cal M}$ of topology $S_2 \times S_2$ with the metric $ds^2 = R_1^2
d\Omega_1^2 + R_2^2 d\Omega_2^2$, where $R_1, R_2 =$const and $d\Omega_a^2 =
d\theta_a^2 + sin^2 \theta_a d\varphi_a^2$.  A possible slicing of ${\cal M}$
can be obtained
by settling $\theta_1 =$const.  This gives 3-manifolds of topology $S_1 \times
S_2$, where $S_1$ is a circle of radius $0 \le r \le R_1$.  One could have
thought that
the configuration with $r=0$ belongs to the regular
boundary.
However, it is easily understood that such configurations cannot be obtained as
critical slices using a Morse function.  In order to give slices of $\theta_1
=$const, the Morse function should be a function only of $\theta_1$, but then
$det (\partial_\mu \partial_\nu f) =0$, and the critical points are always
degenerate.  To see what is wrong with degenerate critical points, consider a
doughnut (torus) lying on a horizontal surface and imagine slicing it with
horizontal planes.  The slice at the bottom is a circle.  But if the torus is
slightly tilted, the circular slice disappears, and the bottom slice is an
isolated point.  The circular slice is degenerate in the sense that it is
present only for a very special slicing.  It can be shown that critical points
of a Morse function are always isolated points.

\refis{louvach} J. Louko and T. Vachaspati, Phys. Lett. {\bf B223}, 21 (1989)

\refis{hallou90} J. J. Halliwell and J. Louko, Phys. Rev. {\bf D42}, 3997
(1990)

\refis{ugl} J. Uglum, Phys. Rev. {\bf D46}, 4365 (1992)

\refis{avbook} A. Vilenkin and E. P. S. Shellard "Cosmic strings and other
topological defects", Cambridge University Press, Cambridge (1994)

\refis{foot7} It is interesting to note that discontinuities
resulting from string reconnection are preserved
at later times by the classical string evolution.  They are known as "kinks"
and propagate around the string at the speed of light.

\refis{geroch} R. Geroch, J. Math. Phys. {\bf 8}, 782 (1967)

\refis{genus} Two-dimensional closed manifolds split into two cobordism
classes,
corresponding to even and odd Euler characteristics, respectively.  The Euler
characteristic for a sphere is $E = 2$, for a torus is $E = 0$, and each
additional handle reduces it by $2$.  An example of a manifold with an odd
Euler characteristic is a sphere with opposite points identified ($E = 1$).
Since $2$-manifolds belonging to different cobordism classes cannot be
connected by an interpolating $3$-manifold, the transition amplitude between
them is zero.  One can therefore consistently assume that only manifolds with
even Euler characteristic are included in superspace, as I did in the text.

\refis{horow} G. Horowitz, Class. Quant. Grav. {\bf 8}, 587 (1991)

\refis{hen} M. Henneaux and C. Teitelboim, [Ann. Phys. {\bf 143}, 127 (1982)]
have shown that a consistent quantum theory of a particle in an external field
can be constructed using the particle's proper time as a time coordinate. It
would be interesting to extend this approach to interacting particles. The
interaction vertices would then correspond to topology changing events in
quantum cosmology.

\refis{farhi} See also E.Farhi, Phys. Lett. {\bf B219}, 403 (1989)

\refis{espace} This target space is what Banks called E-space in
Ref.\cite{banks2}

\refis{banks2} T.Banks, in Physicalia Magazine, vol.12, Special issue in honor
of the 60th birthday of R. Brout (Ghent, 1990)

\endreferences

\head{Figure Captions}

{\bf Fig}.{\bf 1}  The probability flow in the minisuperspace model (5.1).
In this
conformal diagram, $\alpha$ plays the role of time and $\phi$ the role of a
spatial coordinate.

{\bf Fig}.{\bf 2}  Topology change in two dimensions.  The initial
configuration
$\Sigma_1$ has topology $S_1$ and the final configuration $\Sigma_2$ has
topology $S_1 \oplus S_1$.  The manifold $\cal M$ interpolating between
$\Sigma_1$ and $\Sigma_2$ is shown embedded in three-dimensional space.  The
Morse function $f(x)$ is given by the projection on the vertical axis.  The
critical point $P$ and the critical section $f(x) = f(P)$ are indicated.

{\bf Fig}.{\bf 3}  Creation of a two-dimensional universe from nothing.  Here,
the
manifold $\cal M$ has a single boundary $\Sigma$, and the critical section
consists of a single point $P$.

{\bf Fig}.{\bf 4}  A loop of string intersects itself and splits into two.
Sharp
angles (kinks) formed at the point of reconnection propagate around the
"daughter" loops at the speed of light.

{\bf Fig}.{\bf 5}  This $\phi^3$ interaction diagram corresponds to two scalar
particles merging into one.

{\bf Fig}.{\bf 6}  Feynman's picture of pair creation in external field.  A
particle
travelling backwards in time from $t = +\infty$ turns around and travels back
to $t = +\infty$.

\endjnl
\end